\documentclass[12pt,apj]{emulateapj}

\usepackage{subfigure}
\usepackage[usenames]{color}

\begin{document}

\shorttitle{On the Misalignment between Chromospheric Features and the Magnetic Field on the Sun}
\shortauthors{Mart\'inez-Sykora et al.}
\title{On the Misalignment between Chromospheric Features and the Magnetic Field on the Sun}

\author{Juan Mart\'inez-Sykora $^{1,2}$}
\email{j.m.sykora@astro.uio.no}
\author{Bart De Pontieu $^{2,3}$}
\author{Mats Carlsson $^{3}$}
\author{Viggo Hansteen $^{3,2}$}

\affil{$^1$ Bay Area Environmental Research Institute, Sonoma, CA 94952, USA}
\affil{$^2$ Lockheed Martin Solar and Astrophysics Laboratory, Palo Alto, CA 94304, USA}
\affil{$^3$ Institute of Theoretical Astrophysics, University of Oslo, P.O. Box 1029 Blindern, N-0315 Oslo, Norway}

\newcommand{\eg}{{\it e.g.,}} 
\newcommand{\myemail}{juanms@astro.uio.no}
\newcommand{\komment}[1]{\texttt{#1}}

\begin{abstract}
Observations of the upper chromosphere shows an enormous
amount of intricate fine structure. Much of this comes in the form of
linear features which are most often assumed to be well aligned with
the direction of the magnetic field in the low plasma $\beta$ regime
thought to dominate the upper chromosphere.
We use advanced radiative MHD simulations including the
effects of ion-neutral interactions (using the generalized Ohm's law)
in the partially ionized chromosphere to show that the magnetic field
is often not well aligned with chromospheric features. This occurs
where the ambipolar diffusion is large, i.e., ions and neutral populations decouple as the
ion-neutral collision frequency drops, allowing the field to slip through the neutral population,
currents perpendicular to the field are strong, and thermodynamic
timescales are longer than or similar to
those of ambipolar diffusion. We find this often happens in
dynamic spicule or fibril-like features at the top of the
chromosphere. This has important consequences for field extrapolation methods which increasingly use such upper
chromospheric features to help constrain the chromospheric magnetic
field: our results invalidate the underlying
assumption that these features are aligned with the field. 
In addition, our results cast doubt on results from 1D hydrodynamic
models, which assume that plasma remains on the same field lines. Finally,
our simulations show that ambipolar diffusion significantly alters the
amount of free energy available in the coronal part of our simulated
volume, which is likely to have consequences for studies of flare initiation. 

\end{abstract}


\section{Introduction}
Optically thick chromospheric spectral lines such as \ion{Ca}{2}
8542\AA\ or H$\alpha$ are formed over a wide range of
heights from the photospheric line wings to the middle or upper
chromosphere line core. Observations in these lines show a dramatic
transition from wing features that appear to be dominated by
convective motions or acoustic waves in the high plasma $\beta$ regime
(gas pressure is larger than magnetic pressure), to more linear features in the
core of the lines that appear to trace magnetic field lines
\citep[e.g.][]{luc2007,Cauzzi:2008jk}.

Because these linear features are most often assumed to reveal the direction of the
magnetic field, chromospheric structuring is increasingly being used
to help constrain magnetic field extrapolation codes
\citep[e.g.][]{Wiegelmann:2008ud,Jing:2011qa,Aschwanden:2016fj,Aschwanden:2016qy,Zhu:2016fk}. 
Such codes typically use non-linear force-free
field extrapolation methods based on photospheric magnetic field
measurements. Since the boundary conditions are most
readily measured in the photosphere but the field is not necessarily
of a force-free nature at those heights, various
methods are used to preprocess the magnetic field measurements
\citep[e.g.][]{Regnier:2013nx} or to incorporate magnetic field
information from a more force-free boundary region such as the upper
chromosphere \citep{Metcalf:2008bd}. The latter method is entirely
dependent on the assumption that chromospheric features such as
fibrils and spicules, which dominate the upper chromosphere,
are well aligned with the magnetic field. For example,
\citet{Aschwanden:2016qy} extrapolated magnetic field lines using
photospheric vector field measurements from the 
Helioseismic Magnetic Imager \citep[HMI,][]{Scherrer:2012qf} onboard 
of the Solar Dynamic Observatory \citep[SDO,][]{Pesnell:2012nr} using  
the Vertical-Current Approximation Non-linear Force Free Field code 
\citep[VAC-NLFFF code,][]{Aschwanden:2016fj} which finds the best match of the extrapolated 
field lines with chromospheric and coronal features observed with the following imaging instruments: the Interferometric 
Bidimensional Spectrometre \citep[IBIS,][]{Cavallini:2006yq}, the Rapid Oscillation in the Solar 
Atmosphere instrument \citep[ROSA,][]{Jess:2010rt}, the Interface Region Imaging 
Spectrograph \citep[IRIS,][]{De-Pontieu:2014vn}, and Atmospheric Imager Assembly 
\citep[AIA,][]{Lemen:2012uq} onboard of SDO. 

Are these chromospheric structures really aligned with magnetic field
lines? In the single-fluid MHD framework, the assumption that the upper
chromosphere in the vicinity of network or plage is in the low plasma
$\beta$ regime and the observed features thus aligned with the
magnetic field, seems reasonable. However, there are observational
clues that this may not always be the case. For example, \citet{de-la-Cruz-Rodriguez:2011qd} show 
that in some cases chromospheric fibrils do not necessarily follow the magnetic field 
structures. They used Stokes profile observations of \ion{Ca}{2} 8542~\AA\ from 
the Spectro-Polarimeter for INfrared and Optical Regions 
\citep[SPINOR,][]{Socas-Navarro:2006ul}  at the Dunn Solar 
Telescope and CRisp Imaging Spectro-Polarimeter \citep[CRISP,][]{Scharmer:2006gf}
in full Stokes mode at the Swedish 1-m Solar Telescope
\cite[SST,][]{Scharmer:2003ve} and found that the misalignment of some
fibrils with the magnetic field lines can be larger than 45 degrees.  

What causes such a large deviation from the magnetic field direction?
It is already known that, in principle, the neutral population in the partially ionized
chromosphere can have an impact on the force-free nature of the
chromospheric field \citep{Arber:2009ve}. Is it possible that neutrals
can also lead to misalignment of chromospheric features and the
magnetic field? 

In order to better understand this misalignment of the magnetic field lines
with upper chromospheric features, we performed 2D advanced radiative MHD simulations using 
the {\it Bifrost} code \citep{Gudiksen:2011qy}. We included the effects of
the interaction between ions and neutrals in the magnetized and
partially ionized gas of the middle to upper chromosphere by including
ambipolar diffusion in the induction equation (the so-called
generalized Ohm's law) of our MHD code. Our simulations show that while neutrals are mostly coupled
to the magnetic field through collisions with ions, under certain
conditions the collisional frequency between neutrals and ions is low enough
that the ions can become somewhat decoupled from the magnetic field,
thereby allowing the magnetic field to diffuse and magnetic energy to
be dissipated into thermal energy \citep[see][among others]{cowling1957,Braginskii:1965ul,Parker:2007lr}.

We find that our simulation naturally produces 
misalignment of the magnetic field and spicules. 
Our model thus provides an explanation for why some of the observations
show field lines that are misaligned with chromospheric features, and
which conditions can lead to such misalignment. 

\section{Model description}~\label{sec:equations}

The {\it Bifrost} code solves the full MHD equations 
with non-grey, non-LTE radiative transfer \citep{Hayek:2010ac,Carlsson:2012uq} 
and thermal conduction along the magnetic field. The code is described in detail 
in \citet{Gudiksen:2011qy}. In addition, we have also included 
ion-neutral interaction effects adding two 
new terms to the induction equation, i.e., the Hall term and the ambipolar diffusion:

\begin{eqnarray}
\frac{\partial {\bf B}}{\partial t} = && \nabla \times [{\bf u \times B} -  \eta {\bf J}
 - \frac{\eta_{\rm Hall}}{ |B|} {\bf J \times B} +  \nonumber \\ 
&& \frac{\eta_{\rm amb}}{ B^2} ({\bf J \times B}) \times {\bf B} ]\label{eq:faradtot2}
\end{eqnarray}

\noindent
where ${\bf B}$, ${\bf J}$, ${\bf u}$, and $\eta$, $\eta_{\rm Hall}$, $\eta_{\rm amb}$ are the magnetic 
field, the current density, velocity field, the ohmic diffusion, the Hall term, and the ambipolar 
diffusion, respectively \citep[see][ for the derivation of this 
equation]{cowling1957,Braginskii:1965ul}.
\citet{Martinez-Sykora:2016a} describes the details of the implementation of the 
Hall term and ambipolar diffusion in the  {\it Bifrost} code and
extensive tests of this code are described by \citet{Martinez-Sykora:2012uq}. 
The collision cross sections used here are the measurements and calculations listed by 
\citet{Vranjes:2013ve}. 

For this work, it is relevant to reformulate expression~\ref{eq:faradtot2} as follows:
 
\begin{eqnarray}
\frac{\partial {\bf B}}{\partial t} =  \nabla \times \left[{\bf u \times B} - \eta {\bf J}
 - {\bf u_H \times B} +{\bf u_A \times B}\right]\label{eq:faradtot3}
\end{eqnarray}

\noindent where the Hall {\it velocity} is ${\bf u_H}=(\eta_{\rm Hall}{\bf J})/|B| $ 
and the ambipolar  {\it velocity} is ${\bf u_A}=(\eta_{\rm amb}{\bf J \times B})/B^2 $
\citep[see also][]{Cheung:2012uq,Martinez-Sykora:2016a}. The ambipolar
{\it velocity} can be understood as the velocity drift between ions and 
neutrals, i.e., the magnetic field lines are attached only to the ions. 
To gain a better intuitive grasp of how these effects (and {\it velocities}) act on the
magnetic field, let us imagine a curved magnetic ``field line'' contained in a plane. Since the Hall velocity is a 
function of ${\bf J}$, the Hall velocity moves the magnetic field 
line out of the plane. Because the ambipolar diffusion is a function of 
${\bf J \times B}$, the ambipolar velocity will be perpendicular to the field line but contained
within the plane. Therefore, the ambipolar diffusion relaxes the
magnetic tension of the field line. 

\section{Models and initial conditions}~\label{sec:models}

\begin{figure*}
  \includegraphics[width=0.99\textwidth]{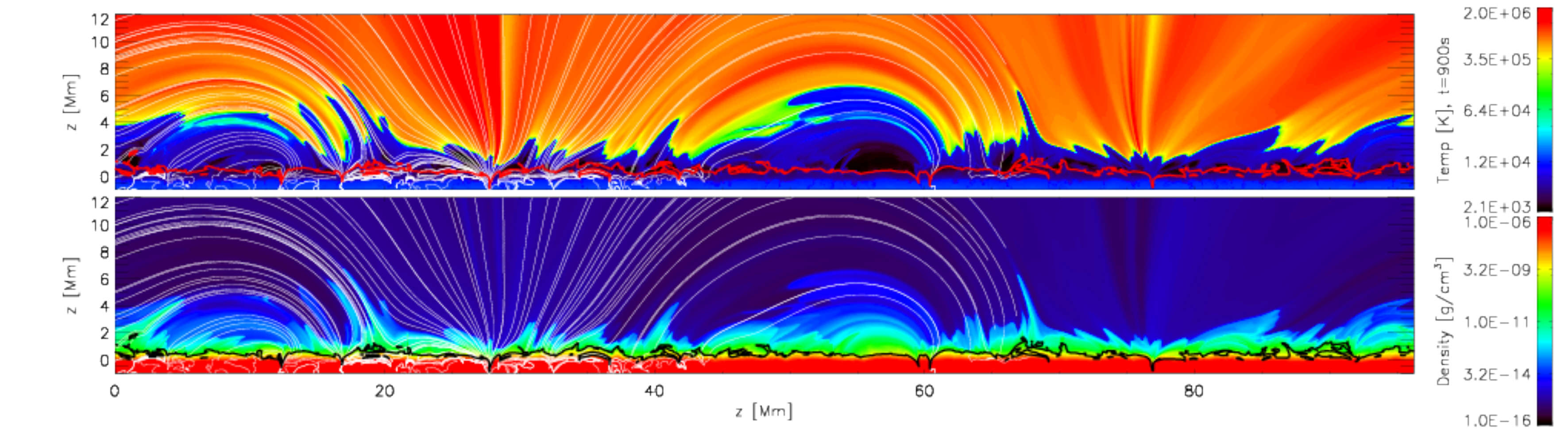} 
 \caption{\label{fig:2dtemp} Temperature map (top panel) and density map 
 (bottom panel) are shown in logarithmic scale. 
 Magnetic field lines are drawn in the left hand side of the maps and plasma 
 beta unity is the thick contour (red in the top panel and black in the bottom panel). }
\end{figure*}

The simulation used here is the so-called GOL (Generalized Ohm's Law)
simulation in \citet{Martinez-Sykora:2016a}, i.e., it is 
a 2.5D model which spans 
from the upper layers of the convection zone ($2.5$~Mm below the 
photosphere) to the corona ($40$~Mm above the photosphere). 
Convective motions perform work on the magnetic field and 
introduce magnetic field stresses in the corona. This energy is dissipated and creates 
the corona self-consistently as the energy deposited by Joule heating is spread 
through thermal conduction \citep{Gudiksen+Nordlund2002} and the temperature 
reaches up to two million degrees (see top panel in Figure~\ref{fig:2dtemp}). The horizontal domain 
spans $96$~Mm. The spatial resolution is uniform along the horizontal axis ($14$~km) 
and non-uniform in the vertical axis allowing smaller grid size where
needed, i.e., in locations such as the photosphere and the transition region ($\sim 12$~km). 

The initial magnetic field has two plage regions of opposite polarity that are connected and 
form loops that are up to $\sim50$~Mm long (Figure~\ref{fig:2dtemp}). 
The mean unsigned field strength at the photosphere is $\sim190$~G. 
The initial magnetic field is a potential field. First we run this setup
for roughly 1.5 hours. After transients have passed through 
the domain, we continue the simulation for another half hour of 
solar time \citep[see][for a detailed description 
of the setup of the model]{Martinez-Sykora:2016a}. 

\section{Results} \label{sec:res}

The ion-neutral interaction effects implemented through the 
generalized Ohm's law, strongly influence the state of the simulated 
chromosphere \citep{Martinez-Sykora:2016a}. Here, we focus on misalignment 
of the magnetic field direction from the chromospheric thermodynamic features.  
 
\begin{figure}
  \includegraphics[width=0.49\textwidth]{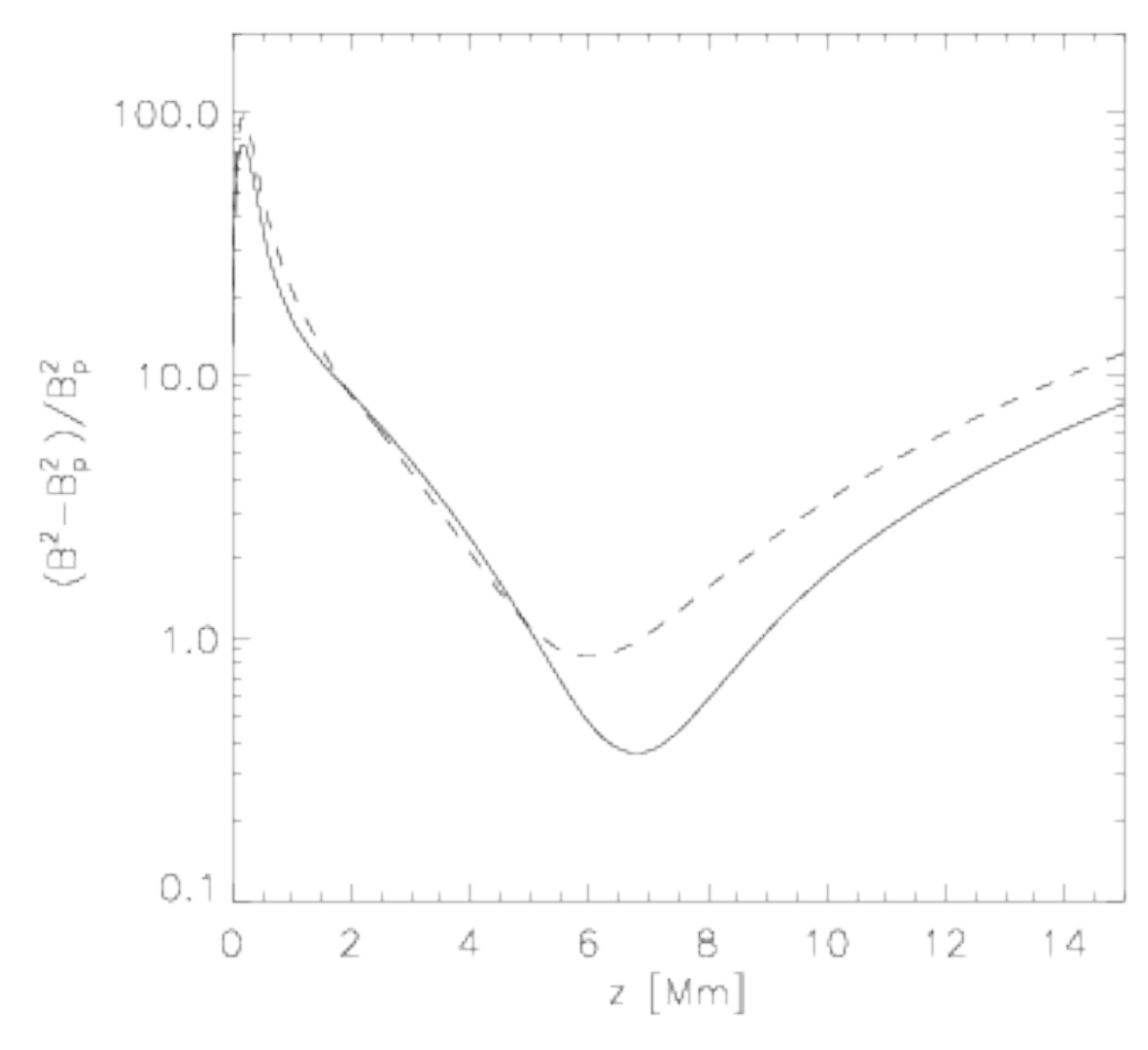}  
  \caption{\label{fig:freeener} Magnetic free energy, normalized by the 
magnetic energy of the potential field extrapolated from z=0~Mm and
integrated over 20 minutes, for the 
simulation that includes ion-neutral interaction effects (solid) and for the equivalent 
simulation which does not include ion-neutral interaction effects (dashed).}
\end{figure}

Generally, the ion-neutral interaction effects impact the magnetic field distribution and 
configuration. In Figure~\ref{fig:freeener} we compare the magnetic free energy from 
the model that includes ion-neutral interaction effects with the free energy in
an equivalent model which has been 
run without ion-neutral interaction effects. The ambipolar diffusion helps to 
accumulate a bit of extra magnetic free energy in the middle-upper chromosphere and transition
region but reduces it in photosphere (~10\%) and a factor between 1.5 and 4 in the corona. 
The magnetic free energy is reduced in the photosphere 
because a small portion of the accumulated magnetic 
field in the photosphere 
is sporadically diffused into the chromosphere in regions 
where the ambipolar diffusion is large enough at low enough heights 
\citep{Martinez-Sykora:2016a}. 
Ion-neutral interaction effects prevent magnetic 
free energy from reaching the corona because it is largely dissipated in 
the chromosphere due to the ambipolar diffusion. 

\begin{figure}
  \includegraphics[width=0.49\textwidth]{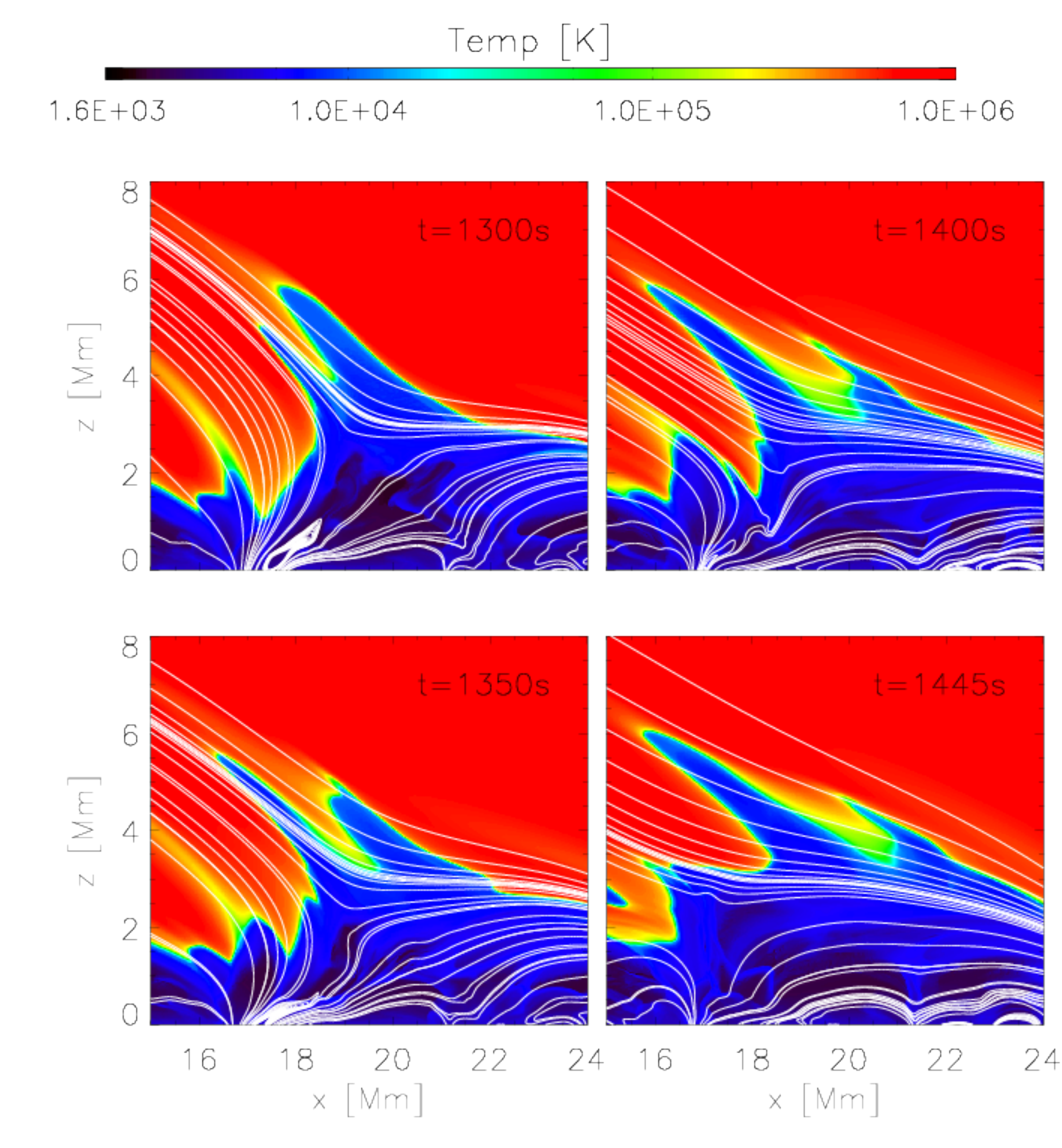}
  \caption{\label{fig:fields} Temperature maps in logarithmic scale at 
  t=1260s (top-left), 1350s (bottom-left), 1445s (top-right) and 1540s (bottom-right) with
  magnetic field lines shown in white. At the beginning (left panels), the thermodynamic structures
  are aligned with the magnetic field, whereas at later times (right panels) the magnetic connectivity
  has changed and the alignment is poor. }
\end{figure}

We also find that, under certain chromospheric conditions, thermodynamic structures may decouple from the magnetic field lines 
due to the ambipolar  {\it velocity}. As a result of this, the simulation shows some spicules and other 
chromospheric features misaligned from the magnetic field during their 
evolution. Figure~\ref{fig:fields} shows temperature maps of the
evolution of several spicules 
with magnetic field lines overplotted as white lines. 
At the early stages of 
spicule evolution, they follow the magnetic field lines. However, as the spicules evolve in time, 
field lines start to decouple from the apparent spicular structure. This figure shows 
misalignment angles of up to \~25 degrees, and this
simulation shows features that can reach a misalignment with 
field lines of up to 40 degrees in the most extreme cases. As a result, 
sometimes, the magnetic field lines 
do not follow the thermal structures in the upper chromosphere and transition region. 

This also impacts the evolution of the features. Towards the end of the evolution
of these spicules, they become wider and their footpoints drift from left to right. 

The thermodynamic chromospheric features do not follow the magnetic 
field lines when the following conditions are met:
1) ambipolar diffusion, magnetic field strength and the current perpendicular to the magnetic 
field are high enough;
2) the timescales of the thermodynamic processes 
are longer than, or of the same order as the ambipolar timescales. 

\begin{figure}
  \includegraphics[width=0.49\textwidth]{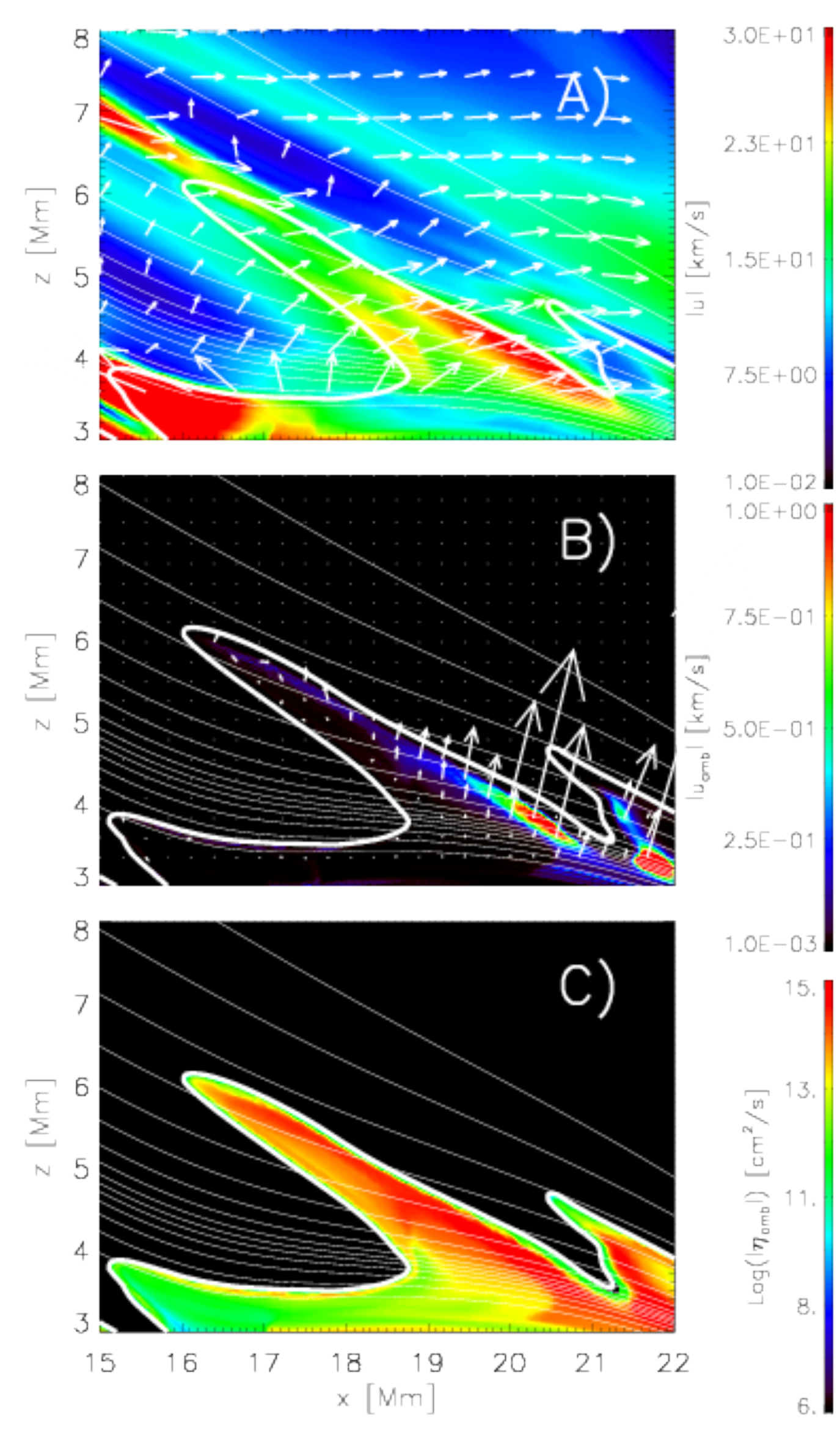}
  \caption{\label{fig:flows} Absolute velocity map with velocity field as white vectors
  are shown in panel A, absolute ambipolar  {\it velocity} map with ambipolar  {\it velocity} field as white vectors
  are shown in panel B, and the ambipolar diffusion in logarithmic scale is shown in panel C. 
  Magnetic field lines are shown as thin white lines and the temperature contour at $10^5$~K
  is shown with a thick white line.}
\end{figure}

This often occurs in upper chromospheric spicules for several reasons.
Since the density drops drastically as a function of height and also as a
function of time (because of the expansion of the spicule), the ambipolar diffusion increases drastically 
(panel C, Figure~\ref{fig:flows}).  
This is a result of low temperature and low ion-neutral collision
frequency. 

One component 
of the magnetic field {\it advection} comes from the ambipolar 
velocity (Panel B) which differs from the advection flows and the direction of 
the spicule and separate the magnetic field lines from the spicule. 
Consequently, the field lines along the spicule have 
two velocity components: 1) one from the advection which is the same as the plasma motion (Equation~\ref{eq:faradtot3}), 
i.e., this flow will move the field lines
and the chromospheric features in the same direction (see the flow 
velocity in Panel A in Figure~\ref{fig:flows}), 2) and the second 
component which is completely detached
from the plasma motion and perpendicular to the field lines 
(see the ambipolar  {\it velocity} in Panel B in Figure~\ref{fig:flows}). 
The latter component is the one that leads to a misalignment 
of the field lines with the spicules. This misalignment is large 
as long as the displacement of the field lines due to the ambipolar 
{\it velocity} is large enough during the lifetime of the spicule. 

\section{Discussion} \label{sec:con}

Our 2.5D radiative MHD simulation includes ion-neutral interaction 
effects and produces some examples of chromospheric features that are
decoupled from the magnetic field direction. This is a result of
ion-neutral interaction effects in the chromosphere and can occur
when the ambipolar diffusion and 
the current perpendicular to the magnetic field lines are large, and the 
thermo-dynamic timescales are at least of the same order as the 
ambipolar velocity timescales. The simulated features that become misaligned from the 
magnetic field have lifetimes of order a few minutes. The time-scale becomes shorter in 
regions with large currents perpendicular to the magnetic field lines and 
large ambipolar diffusion due to low values of the temperature, ion-neutral collision 
frequency and/or ionization degree. 
Under these conditions the magnetic field may undergo evolution that 
is different from that of the thermodynamic 
structures. For example, decaying spicules may change 
their connectivity with ``different'' field lines crossing the
spicule. In such a case, instead of following a space-time parabolic 
profile along a fixed direction, the spicule may show a 
horizontal displacement at the same time as they disappear (towards
the end of their lifetime). This 
process can provide a natural explanation for the observations of 
\citet{de-la-Cruz-Rodriguez:2011qd} where fibrils or spicule-like features do not necessarily 
follow the magnetic field direction. 

Our model shows that the
misalignment is not uniform in space or time, with some structures
less affected. In addition, dynamic features appear to be typically
more misaligned towards the end of their lifetime. This occurs in particular in regions with enough 
current perpendicular to the magnetic field, i.e., where there is 
large magnetic tension. Such conditions can also be expected in active
regions with strong currents, such as in newly emerging active
regions. However, it is a priori not clear how one can determine from
the observations alone which features are likely not well aligned with
the magnetic field. Future work will
be needed to investigate how misalignment can be estimated based on
observational clues, such as the temporal behavior of the structures
or the presence of currents. 

Our results suggest that ion-neutral interaction effects may have a significant impact on 
magnetic field extrapolation methods. Within the chromosphere, the ambipolar
diffusion shows strong variations in both space and time. In regions where the 
ambipolar diffusion is strong the magnetic
field will be more potential. However, at the boundaries between regions of 
strong and weak ambipolar diffusion, the magnetic field lines may have strong 
changes in the connectivity. 
These variations in space and time of the ambipolar diffusion 
impact the magneto-thermodynamic processes in the chromosphere 
\citep{Martinez-Sykora:2016a}. 
This was missing in previous numerical models \citep{Carlsson:2016rt}. 
It is unclear how such spatial and
temporal complexity can be
captured in field extrapolation methods that are based on photospheric vector magnetograms. 

In addition, magnetic field extrapolation codes that attempt to use the direction 
of chromospheric features in order to find the best match with the field lines 
may provide field configurations that are incorrect. This is because the magnetic field  
may not be well aligned with the chromospheric features due to the presence of ambipolar
diffusion.  As a result of this, the measurements of the misalignment between features and
magnetic field extrapolation using these methods may be incorrect and 
in general underestimated \citep{Aschwanden:2016qy}. Future studies of these 
extrapolation codes should investigate this issue further by trying
the method on synthetic data from this type of radiative MHD models
that include ion-neutral interaction effects and comparing with the
actual magnetic field.  

Many studies of the solar atmosphere, in particular the corona, are undertaken using 1D
hydrodynamic loop models \citep[e.g.][]{Klimchuk:2014fk}. Such models
are based on the assumption that the thermodynamic evolution occurs along magnetic field 
lines or tubes (1D). However, our results indicate that decoupling of the field lines from plasma 
is a frequent occurrence in the chromosphere. This will change the
connectivity of each element along the 1D models. In addition, 
ambipolar diffusion fundamentally undermines the assumption that the plasma is tied to the field on 
timescales of many minutes. As we have shown, this is not always the case. Therefore, 
these 1D models cannot capture or mimic the physics of the processes
that connect the corona to the chromosphere and transition region. 

\cite{Leake:2006kx} performed 2D simulations of flux emergence with ambipolar diffusion. They 
also notice that the magnetic field is more potential in the atmosphere due to the ambipolar 
dissipation. However, they get several orders of magnitude smaller currents for the 
case with ambipolar diffusion than without ambipolar diffusion whereas we get 
from 1.5 to 4 times smaller in the case for ambipolar diffusion than without ambipolar 
diffusion. This is most likely due to the highly simplified setup of
the ambipolar diffusion in their model chromosphere and also due to their missing 
many of the chromospheric processes driven by the convective motion. 

Our results also have a potential impact on more advanced 3D radiative
MHD models. This is because the magnetic 
field energy deposition in the corona in radiative MHD models will change as soon as 
ion-neutral interaction effects are introduced. For example, 
\citet{Peter:2015zv} noticed that larger domains
show stronger flows and Doppler shifts for transition region EUV
profiles (similar to observations) compared to smaller simulated
domains 
\citep{Hansteen:2010uq,Hansteen:2015qv}. Ion-neutral interaction
effects will impact these results and deeper investigations that
include ion-neutral effects must be performed for computational domains of various sizes.

Our simulation suffers from several limitations that need to be 
addressed. Our simulation does not include time dependent ionization which will impact the 
spatial and temporal distribution of ambipolar diffusion in the chromosphere \citep{Leenaarts:2007sf,Golding:2014fk}. The process described above is 
also strongly constrained to the two dimensions of the model, therefore the
expansion of these models into three dimensions is needed. 
Finally, the Generalized Ohm's law is valid as long as 
the time scales are much larger than the ion-neutral collision frequencies, but 
in the transition region this may not always be fulfilled and ions may decouple 
from neutrals \citep{Martinez-Sykora:2012uq}. This can potentially alter these results.

\section{Acknowledgments}

We gratefully acknowledge support by NASA grants NNX11AN98G,
NNM12AB40P, NNH15ZDA001N-HSR, NNX16AG90G, 
and NASA contracts NNM07AA01C (Hinode), and NNG09FA40C 
(IRIS). This research was supported by the Research Council of Norway and by the 
European Research Council under the European Union's Seventh Framework 
Programme (FP7/2007-2013) / ERC Grant agreement nr. 291058.
The simulations have been run on clusters from the Notur project, 
and the Pleiades cluster through the computing project s1061 from the High 
End Computing (HEC) division of NASA. We thankfully acknowledge the 
computer and supercomputer resources of the Research Council of Norway 
through grant 170935/V30 and through grants of computing time from the 
Programme for Supercomputing. This work has benefited from discussions at 
the International Space Science Institute (ISSI) meetings on 
``Heating of the magnetized chromosphere'' where many aspects of this 
paper were discussed with other colleagues.
To analyze the data we have used IDL.

\end{document}